\documentstyle[epsfig,12pt]{article}

\begin{document}

\begin{titlepage}

\vspace*{2.truecm}

\centerline{\large \bf  Monte Carlo Simulations of Critical Dynamics }
\vskip 0.6truecm
\centerline{\large \bf with Conserved Order Parameter}
\vskip 0.6truecm

\vskip 2.0truecm
\centerline{\bf B. Zheng}
\vskip 0.2truecm

\vskip 0.2truecm
\centerline{FB Physik, Universit\"at Halle, 06099 Halle, Germany}
\vskip 0.2truecm
\centerline{Institute of Theoretical Physics, Academia Sinica,
      100080 Beijing, P. R. China}
  
\vskip 2.5truecm

\abstract{Taking the two-dimensional Ising model for example,
short-time behavior of critical dynamics with a conserved order
parameter is investigated by Monte Carlo simulations.
Scaling behavior is observed,
but the dynamic exponent $z$ is updating schemes dependent.
}

\vspace{0.5cm}



\end{titlepage}

Traditionally, it is known that there exists
universal dynamic scaling behavior 
in the long-time regime of critical dynamic evolution,
since both the spatial correlation length and the correlation time
are very large.
In recent years, however, it is discovered that 
universal scaling behavior emerges already in the {\it macroscopic
short-time} regime, after a microscopic time scale $t_{mic}$ which is large
enough in the microscopic sense 
\cite {jan89,hus89,sta92,li94,sch95,gra95,zhe98}.
Important is that one should take care of the macroscopic 
initial conditions carefully.

For a {\it random} initial state, for example, 
at the critical point 
the non-equilibrium spatial correlation length increases
in time by a power law $t^{1/z}$ due to 
the {\it divergent} correlation time \cite {zhe98}.
 This implies a kind of memory effect
in the system. Even though the spatial correlation length
($\sim t^{1/z}$) is still small in the short-time regime,
one may expect dynamic scaling.
For a small initial magnetization, an extra critical
exponent is sufficient to describe its
scaling behavior \cite {jan89,sch95,zhe98}.
In general, one needs a characteristic function
for an arbitrary magnetization \cite {zhe96,che00}.
If the initial spatial correlation length 
is non-zero, it induces corrections to scaling \cite {jan89}.

Since 1989, analytical calculations for short-time dynamic scaling behavior
based on renormalization group
methods \cite {jan89,oer93,oer94} 
have extended to different dynamics
classified by Hohenberg and Halperin \cite {hoh77}.
 However, Monte Carlo simulations
are still limited in dynamics of model A 
\cite {zhe98,men98,tom98a,bru99,zhe99b,luo99,zha99,bra00,jen00,yin00}.
Actually, due to severe critical slowing down,
Monte Carlo simulations for long-time behavior of
critical dynamics beyond model A are also in 
 preliminary stage. On the other hand,
short-time dynamic scaling is not only conceptually interesting but
also practically important. It provides new methods
for numerical measurements of not only dynamic exponents but
also static exponents as well as the critical temperature
\cite {li95,luo98}.
These methods do not suffer from critical slowing down.
Therefore it is important to investigate short-time
scaling behavior of critical dynamics
beyond model A with Monte Carlo methods.

Dynamics of model A is a kind of relaxation
dynamics {\it without} relevant conserved 
quantities \cite {hoh77}.
 If we consider only dynamic relaxation 
processes, dynamics with different relevant conserved quantities
is classified into model B, C and D.
In this letter, taking the two-dimensional Ising model for example,
with Monte Carlo methods we study 
dynamics {\it with a conserved order parameter}, which is called model B,
according to Ref. \cite {hoh77}.

In dynamics of model B, to guarantee criticality
the conserved order parameter must be set to zero.
The equilibrium state of model B is in a {\it same} universality
class of model A. For lattice models,
the critical temperature in dynamics of model B is also
the same as that of model A.  
Of course, the dynamic exponent $z$ of model B
is different from that of model A.
For the $\phi^4$ theory with a generalized 
Langevin equation in which the order parameter is conserved,
in the long-time regime of dynamic evolution
it has been derived that $z=4-\eta$, with $\eta$ being the well known 
static exponent \cite {hoh77}. In Ref. \cite {jan89}, it is argued
that there exists also short-time dynamic scaling
and no new independent critical exponents emerge.

For the Ising model, a realization of dynamics of model B is
considered to be Kawasaki dynamics in which two neighboring spins exchange 
their values in a flip. 
Early simulations with small lattices for 
dynamic behavior in {\it equilibrium} yield
a dynamic exponent $z$ consistent with
the theoretical prediction 
$z=4-\eta$ \cite {yal82}. In this paper, we study the 
non-equilibrium dynamic
process starting from a {\it random} initial state.
We measure the spatial correlation function
\begin{equation}
C(r,t)=\frac {1}{L^d} \sum_i \langle S_i (t) S_{i+r} (t)  \rangle\ .
\label{e10}
\end{equation}
Here $S_i$ is a Ising spin, 
$d=2$ is the space dimension and $L$ is the lattice size.
In updating,
 we randomly pick up neighboring spin pairs.
When $L^d$ pairs of spins are updated, we define one 
Monte Carlo time step.
At the critical point, standard scaling form for $C(r,t)$ is written as
\begin{equation}
C(r,t)=t^{-2\beta/\nu z} F(\frac {r}{t^{1/z}}).
\label{e20}
\end{equation}
Here $\beta$ and $\nu$ are static exponents and
$z$ is the dynamic exponent.
Usually such a scaling form is assumed to be valid
already in the macroscopic short-time regime,
after a time scale $t_{mic}$ which is large enough in the
microscopic sense. If a Monte Carlo time step 
is considered as a typical microscopic time unit,
$t_{mic}$ may be several to some hundred time steps.
For dynamics of model A, this is indeed the case
\cite {zhe98}.

\begin{figure}[t]\centering
\epsfysize=10cm
\epsfclipoff
\fboxsep=0pt
\setlength{\unitlength}{1cm}
\begin{picture}(13.6,12)(0,0)
\put(-1,0){{\epsffile{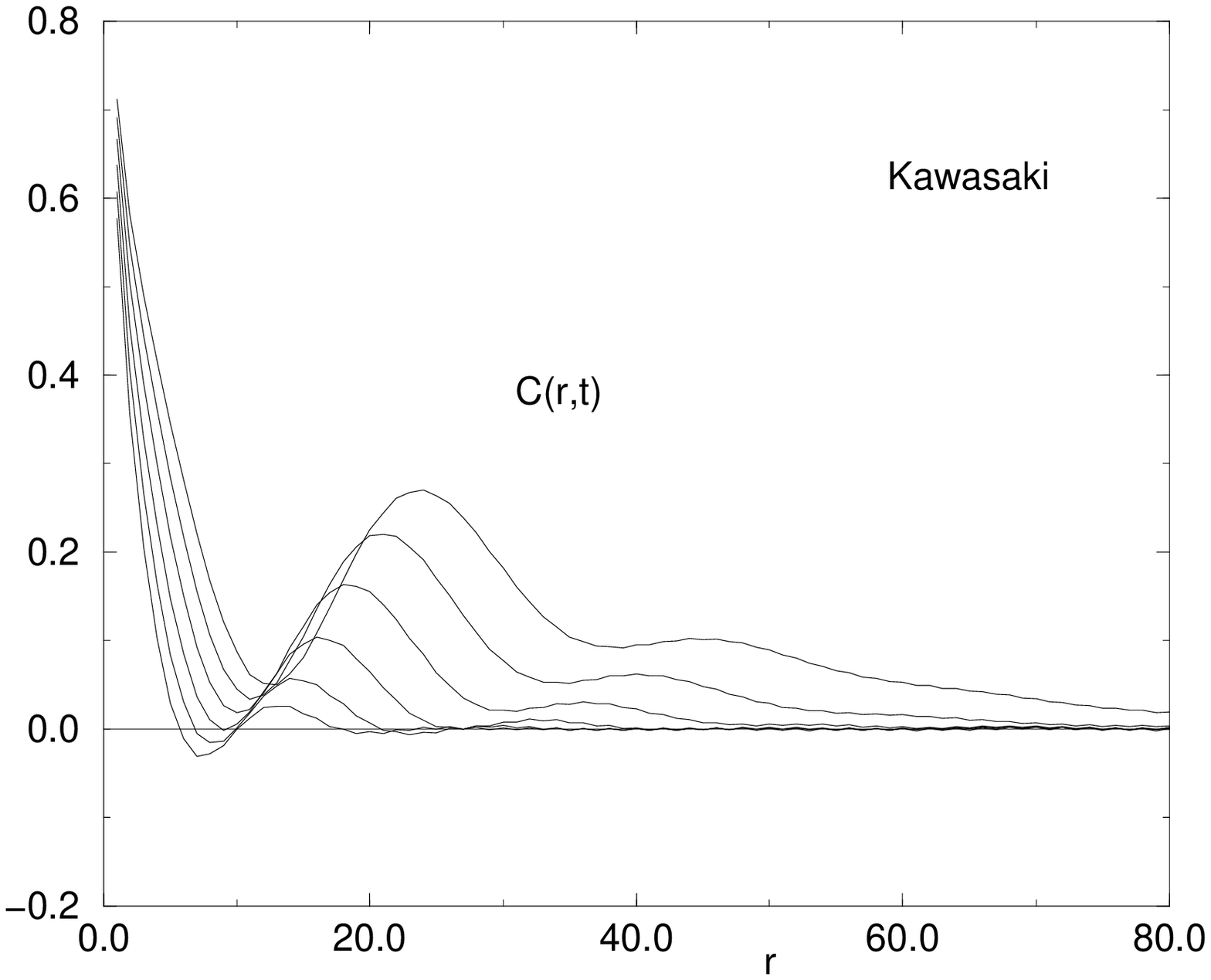}}}
\end{picture}
\caption{Correlation functions $C(r,t)$ for Kawasaki dynamics with 
the heat-bath algorithm. The lattice size is $L=512$.
From left to right, curves correspond to
the time $t=2000$, $4000$, $8000$,
$16000$, $32000$ and $64000$.
}
\label{f1}
\end{figure}

In Fig. \ref {f1}, the spatial correlation function
$C(r,t)$ for Kawasaki dynamics with the heat-bath algorithm is displayed.
The lattice size is $L=512$ and samples of
initial configurations for averaging
are $100$. Maximum updating time is $64000$.
Here finite size effect is apparently negligible.
For example, $C(r,t)$  at $r=80$ is already very small. 
Extra simulations up to $t=256000$ show similar behavior
as in Fig. \ref {f1},
but one needs bigger lattices.

Carefully looking at Fig. \ref {f1}, 
we conclude that there is no scaling behavior
as described by Eq. (\ref {e20}),
even though it seems there is a kind of self-similarity
during the dynamic evolution. Let us concentrate 
our attention on
the first peak in $r$ direction (for a fixed time $t$).
If there would be scaling, the first peak 
for different time $t$'s should overlap
each other after properly rescaling $r$ and $C(r,t)$
according to Eq. (\ref {e20}).
We remind ourselves that the exponent $2 \beta/\nu z$
is positive. Therefore the peak 
for a larger $t$ should be lower than that for 
a smaller $t$. But it is not the case in Fig. \ref {f1}.
This fact indicates that at least for the short-time behavior,
Kawasaki dynamics of the Ising model can not be described
by the generalized Langevin equation 
of model B for the $\phi^4$
theory given in Ref. \cite {hoh77}.

Why there exists no short-time dynamic scaling behavior 
in Kawasaki dynamics is not very clear for us.
What we observe is that in Kawasaki dynamics
the order parameter is conserved very locally by exchanging
two neighboring spins. Spin exchanging takes place 
only along domain walls. Smaller domains can not be so
easily created during time evolution.
As a result, the distribution of domain sizes 
(at a fixed time $t$) is rather sharp.
This is the origin of the oscillating mode
in $C(r,t)$ as shown in Fig. \ref {f1}.
This oscillating mode violates the scaling form.

\begin{figure}[t]
\epsfysize=6.cm
\epsfclipoff
\fboxsep=0pt
\setlength{\unitlength}{0.6cm}
\begin{picture}(13.6,12)(0,0)
\put(-2.,0){{\epsffile{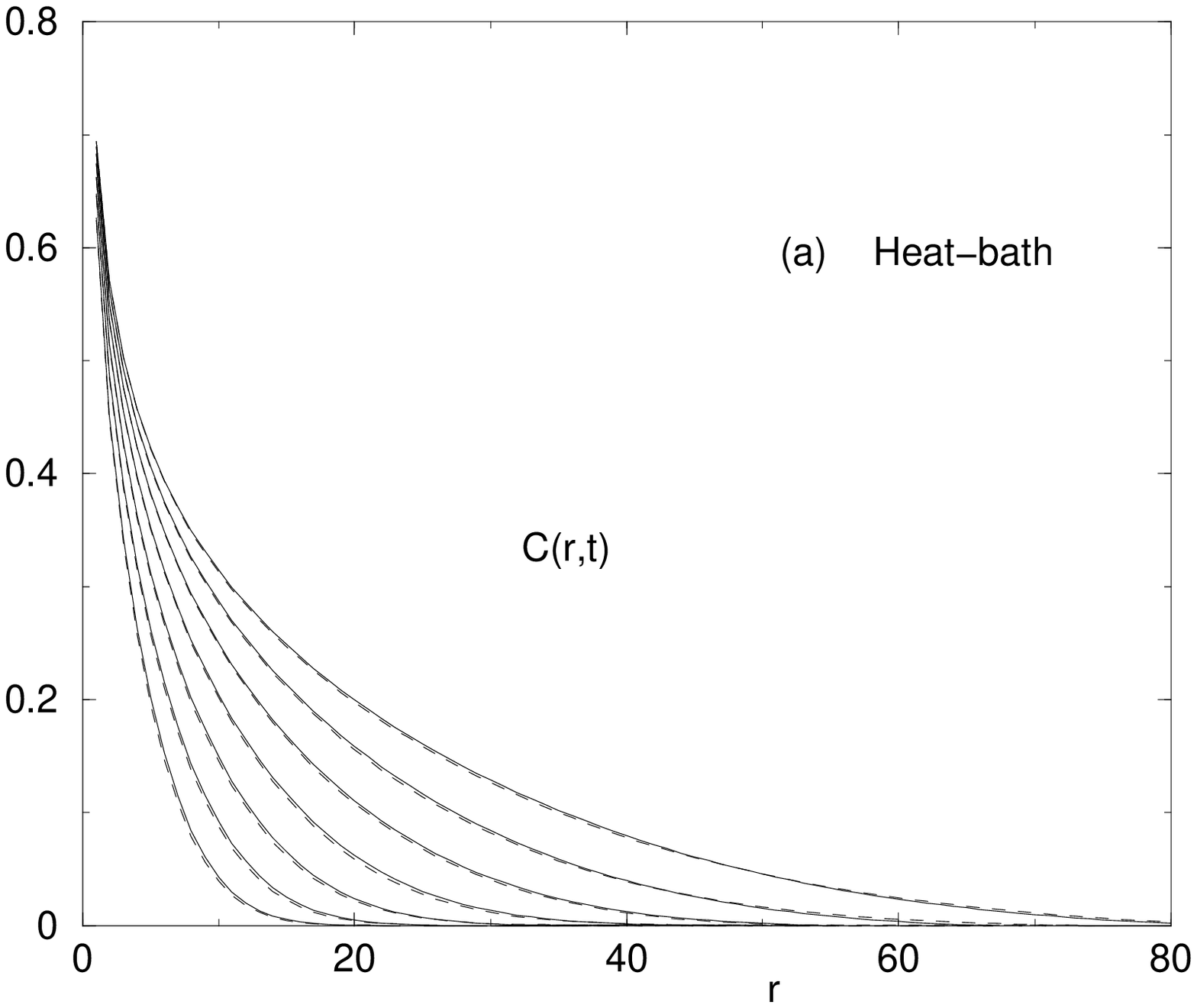}}}
\epsfysize=6.cm
\put(11.5,0){{\epsffile{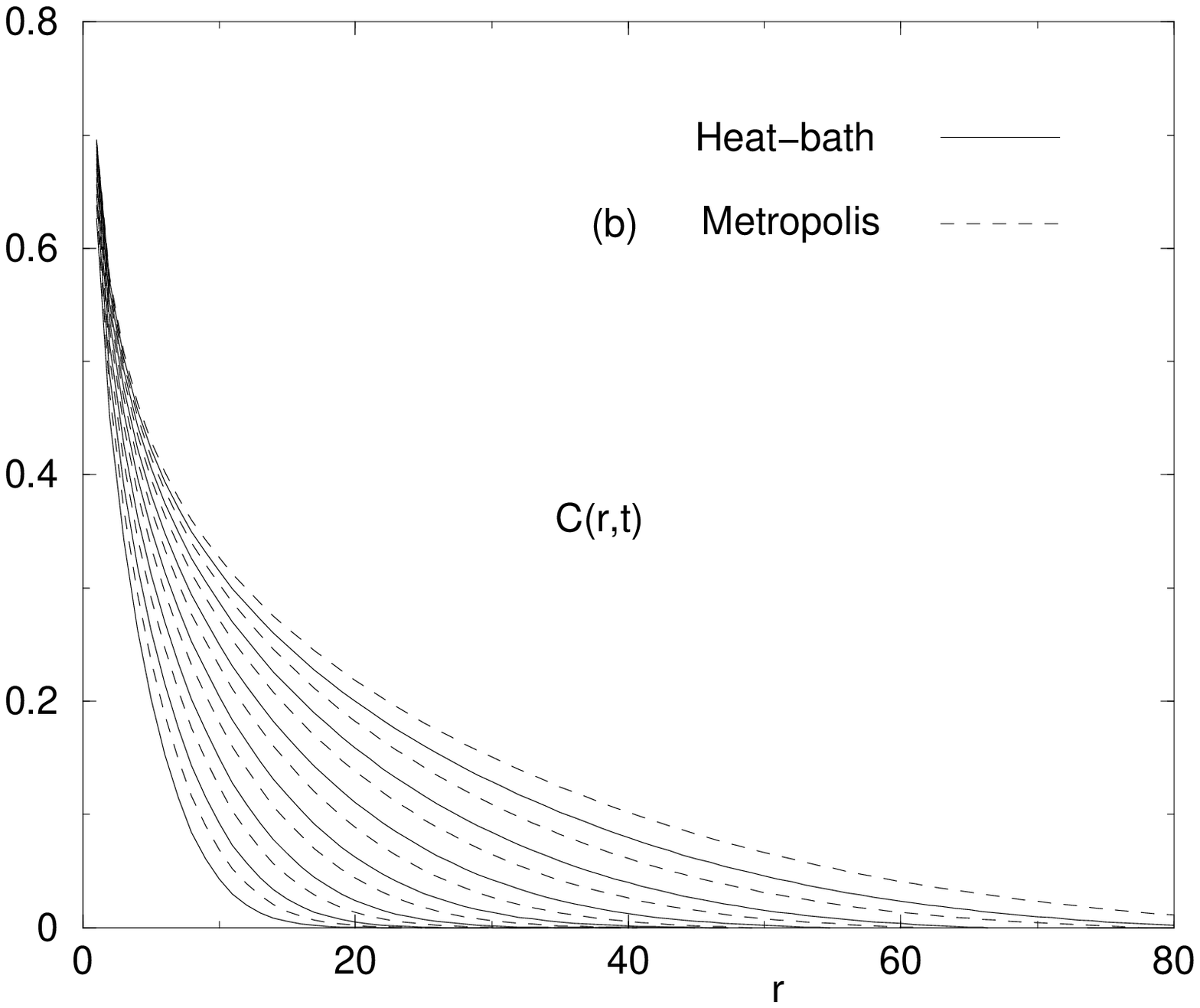}}}
\end{picture}
\caption{Correlation functions $C(r,t)$ for dynamics with
a globally conserved order parameter.
The lattice size is $L=512$.
(a) The heat-bath algorithm. Solid lines are from a updating scheme
in which one spin sweeps over the lattice
regularly. Dashed lines are from a scheme
in which both spins are randomly selected. 
(b) The solid lines are the same as in (a) but dashed lines
are obtained with the Metropolis algorithm.
In both (a) and (b), from left to right, curves 
correspond to the time $t=80$, $160$, $320$,
$640$, $1280$, $2560$ and $5120$.
}
\label{f2}
\end{figure}

Dynamics with a conserved order parameter is
not unique. For example,
we may consider
an alternative dynamic process in which the order parameter
is only {\it globally} conserved. A simple case
is that we exchange two randomly separated spins.
In this case, scaling behavior is observed.
To show the scaling behavior is rather robust,
we have performed the simulations with different
updating schemes and algorithms.
In Fig. \ref {f2} (a), 
correlation functions $C(r,t)$ with the heat-bath algorithm 
are displayed for two different updating schemes.
Solid lines are from a scheme in which one spin 
sweeps over the lattice regularly,
while dashed lines are from a scheme in which both spins are
randomly selected. Total samples of initial configurations
for averaging are $620$ and $220$ respectively.
Obviously, both updating schemes yield
almost the same results, even though some fluctuation is seen
for larger times. 
In Fig. \ref {f2} (b), similar curves are seen for both 
the heat-bath and the Metropolis algorithms.
But Metropolis dynamics evolves slightly faster.
Important in Fig. \ref {f2} is that the oscillating mode
is swept away. Comparing Fig. \ref {f1} and 
\ref {f2}, we observe that Kawasaki dynamics 
is much slower.

\begin{figure}[t]
\epsfysize=6.cm
\epsfclipoff
\fboxsep=0pt
\setlength{\unitlength}{0.6cm}
\begin{picture}(13.6,12)(0,0)
\put(-2.,0){{\epsffile{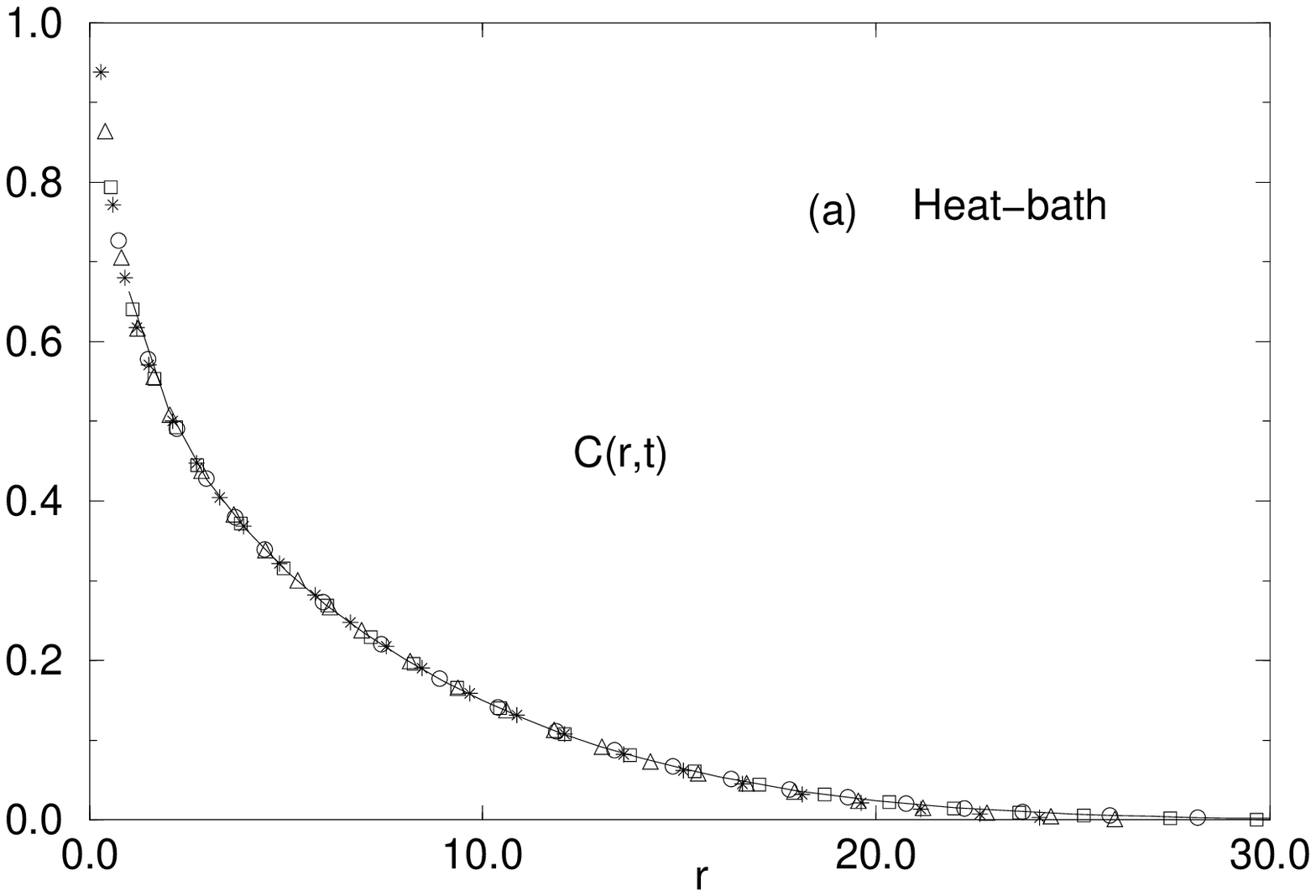}}}
\epsfysize=6.cm
\put(11.5,0){{\epsffile{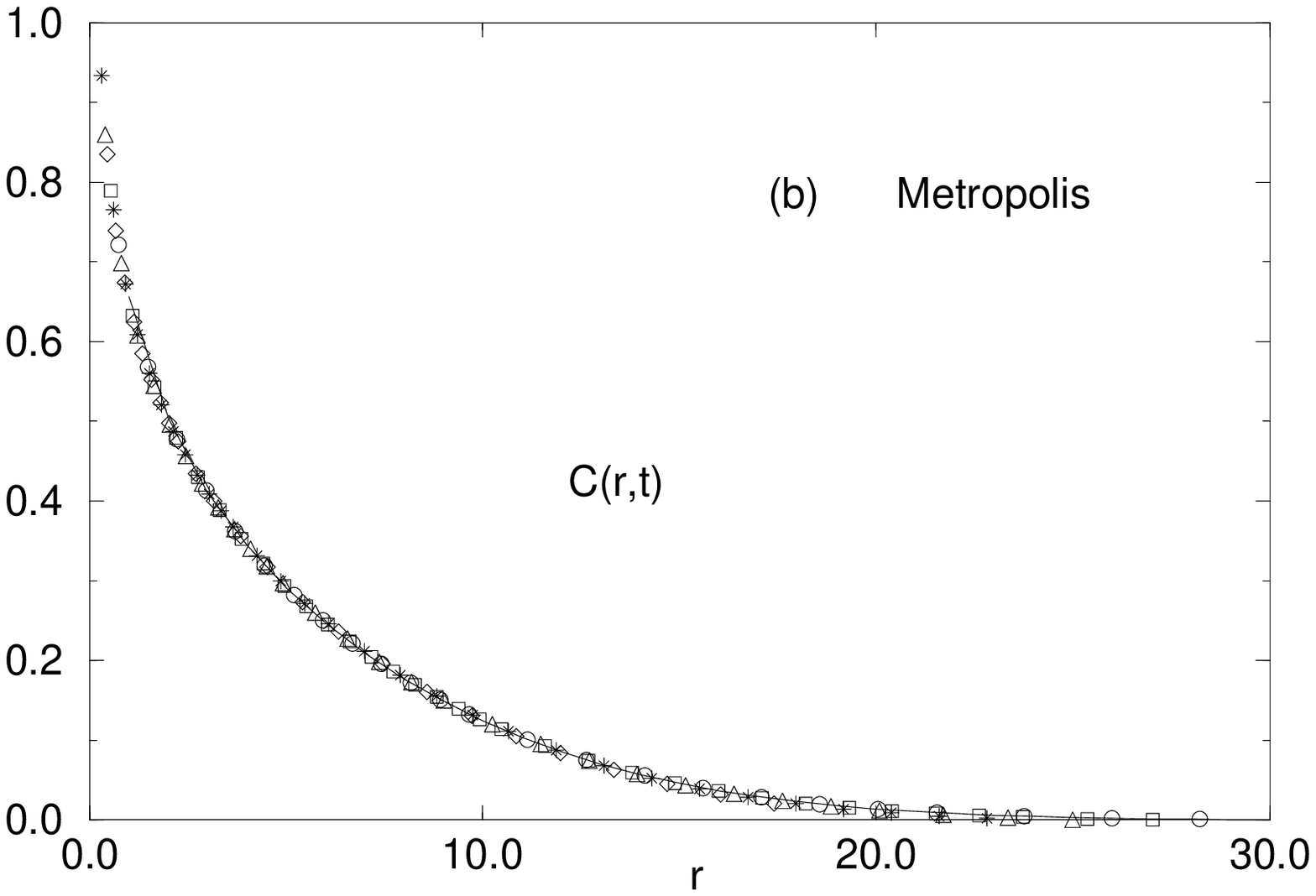}}}
\end{picture}
\caption{Scaling plot for $C(r,t)$.
(a) The solid line is for $t_1=320$, while $\circ$, $\Box$, $\triangle$
and $\ast$ fitted to the curve of $t_1=320$ are those for
$t=640$, $1280$, $2560$ and $5120$ but $r$ and $C$
are rescaled according to
Eq. (\protect {\ref {e20}}) to $r/(t/t_1)^{1/z}$ and 
$(t/t_1)^{2\beta/\nu z}C$ with 
$z=2.315$ and $2\beta/\nu=0.252$. 
(b) The solid line is for $t_1=160$, while
 $\circ$, $\Box$, $\triangle$,
$\ast$ and $\diamond$ are for 
$t=320$, $t=640$, $1280$, $2560$ and $5120$
with $z=2.330$ and $2\beta/\nu=0.251$. 
}
\label{f3}
\end{figure}

Now we study scaling behavior of $C(r,t)$.
According to Eq. (\ref {e20}), if the dynamic exponent $z$ and 
the static exponent $\beta/\nu$ are properly chosen,
$C(r,t)$ at different time $t$'s collapse
on a single curve. In other words, from scaling collapse of $C(r,t)$ 
one can estimate $z$ and $\beta/\nu$.
Very careful analysis shows that 
the microscopic time scale $t_{mic}$'s for the heat -bath
and the Metropolis algorithms are less than $320$ and $160$
respectively. Also, up to the time $t=5120$
there exists no visible finite size effect for both algorithms
with $L=512$.
In Fig. \ref {f3}, scaling plot for $C(r,t)$
is shown. All data points collapse nicely.
The fitted exponents are
$z=2.315(5)$ and $2\beta/\nu=0.252(3)$
for the heat-bath algorithm and
$z=2.330(15)$ and $2\beta/\nu=0.251(2)$
for the Metropolis algorithm.
Results for both algorithms are
well consistent.
The values for the exponent  
$2\beta/\nu$ agree remarkably with
the exact value $0.25$ obtained in equilibrium. 
These results strongly support the scaling form
in Eq. (\ref {e20}). 

If we fix the static exponent
$2\beta/\nu$ to its exact value in the scaling collapse,
the resulting dynamic exponent $z$ is
$z=2.320(5)$ for the heat-bath algorithm
and $z=2.335(10)$ for the Metropolis algorithm.
Taking into account all above measurements,
we conclude 
\begin{equation}
z=2.325(10)\ . 
\end{equation}
This value of $z$ is significantly
different from $z=4-\eta=3.75$. 
In other words, the dynamics is not in a same universality
class of the $\phi^4$ theory with the generalized Langevin
equation in Ref. \cite {hoh77}. 
The value $z=2.325(10)$ is relatively close to 
$z=2.165(10)$ for
dynamics of model A \cite {zhe98},
but also clearly different.

Another interesting dynamic observable is the 
auto-correlation function
\begin{equation}
A(t)=\frac {1}{L^d} \sum_i \langle S_i (0) S_{i} (t)  \rangle\ .
\label{e30}
\end{equation}
In Refs. \cite {jan92,jan89}, it has been derived
that at the critical point, for dynamics of model A, 
$A(t)$ decays by a power law
\begin{equation}
A(t) \sim t^{-\lambda}\ .
\label{e40}
\end{equation}
Here the exponent $\lambda=d/z-(2-z-\eta-\eta_0/2)/z$, and 
$\eta_0$ is an exponent induced from the divergence
related to the initial condition \cite {jan92,jan89}.
For dynamics of model B for the $\phi^4$ theory
discussed in Ref. \cite {hoh77},
it is pointed out in Ref. \cite {jan89},
 $\eta_0=0$, since there is no divergence related
 to the initial condition. 
 
\begin{figure}[t]\centering
\epsfysize=10cm
\epsfclipoff
\fboxsep=0pt
\setlength{\unitlength}{1cm}
\begin{picture}(13.6,12)(0,0)
\put(-1,0){{\epsffile{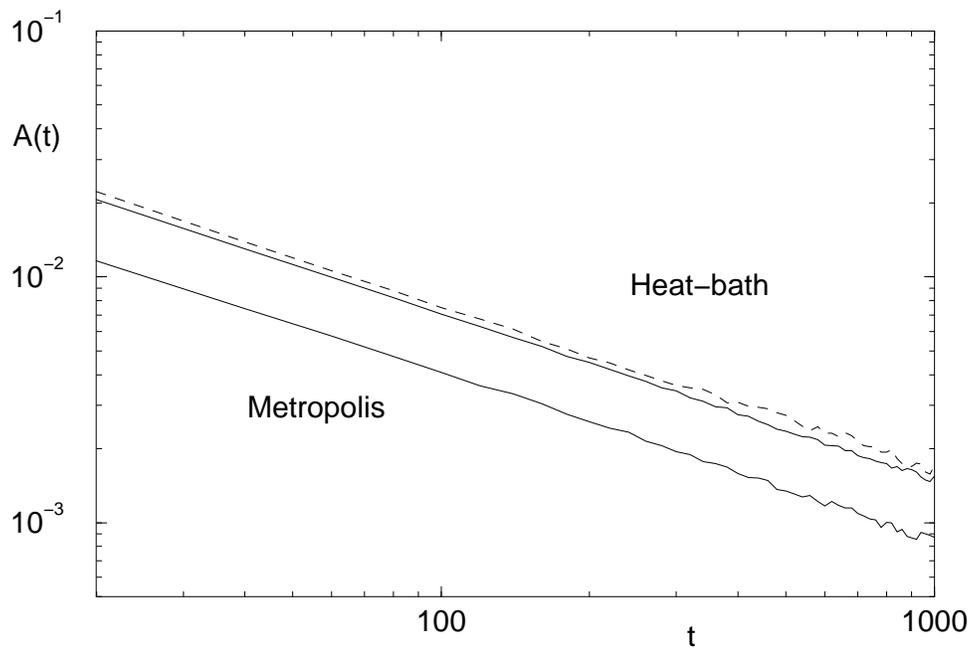}}}
\end{picture}
\caption{Auto-correlation $A(t)$ in double-log scale. 
The dashed line is for the updating scheme with both spins
randomly picked up
}
\label{f4}
\end{figure}

 In Fig. \ref {f4}, the auto-correlation $A(t)$
 in our simulations for Fig. \ref {f2}
 is plotted in double-log scale.
 Power law behavior is observed starting around 
 a time $t=20$. After $t=1000$, fluctuation
 becomes large. In the time interval $[20, 1000]$,
 we measure the exponent $\lambda=0.675(3)$
 and $0.679(13)$ for the heat-bath and the Metropolis algorithms
 respectively. As an average, we estimate
 \begin{equation}
 \lambda=0.677(8)\ .
 \end{equation}
 Obviously, it does not yield $\eta_0=0$
 and also is different from $\lambda=0.737(1)$
 for dynamics of model A \cite {zhe98}.
 
 In conclusions, we have investigated 
 short-time behavior of critical dynamics for the 
 two-dimensional Ising model with a conserved order parameter.
 For a locally conserved order parameter
 (Kawasaki dynamics), standard scaling behavior  
 does not exist. If we release the condition such
 that the order parameter is only globally conserved,
 universal scaling behavior emerges.
 However, the dynamics is not in a same universality class
 of the $\phi^4$ theory
 with the generalized Langevin equation 
which is called model B in Ref. \cite {hoh77}.
Very probably both $z$ and $\lambda$ are independent
critical exponents in this case.
It is actually interesting how to construct
a corresponding continuum model and solve it 
 with renormalization group methods.
It seems that there might be some similarity
at certain points 
 between this dynamics and kinetic roughening
\cite {lop99}. 
\vskip 2.5truecm

{\bf Erratum}

\begin{figure}[t]\centering
\epsfysize=10cm
\epsfclipoff
\fboxsep=0pt
\setlength{\unitlength}{1cm}
\begin{picture}(13.6,12)(0,0)
\put(-1,0){{\epsffile{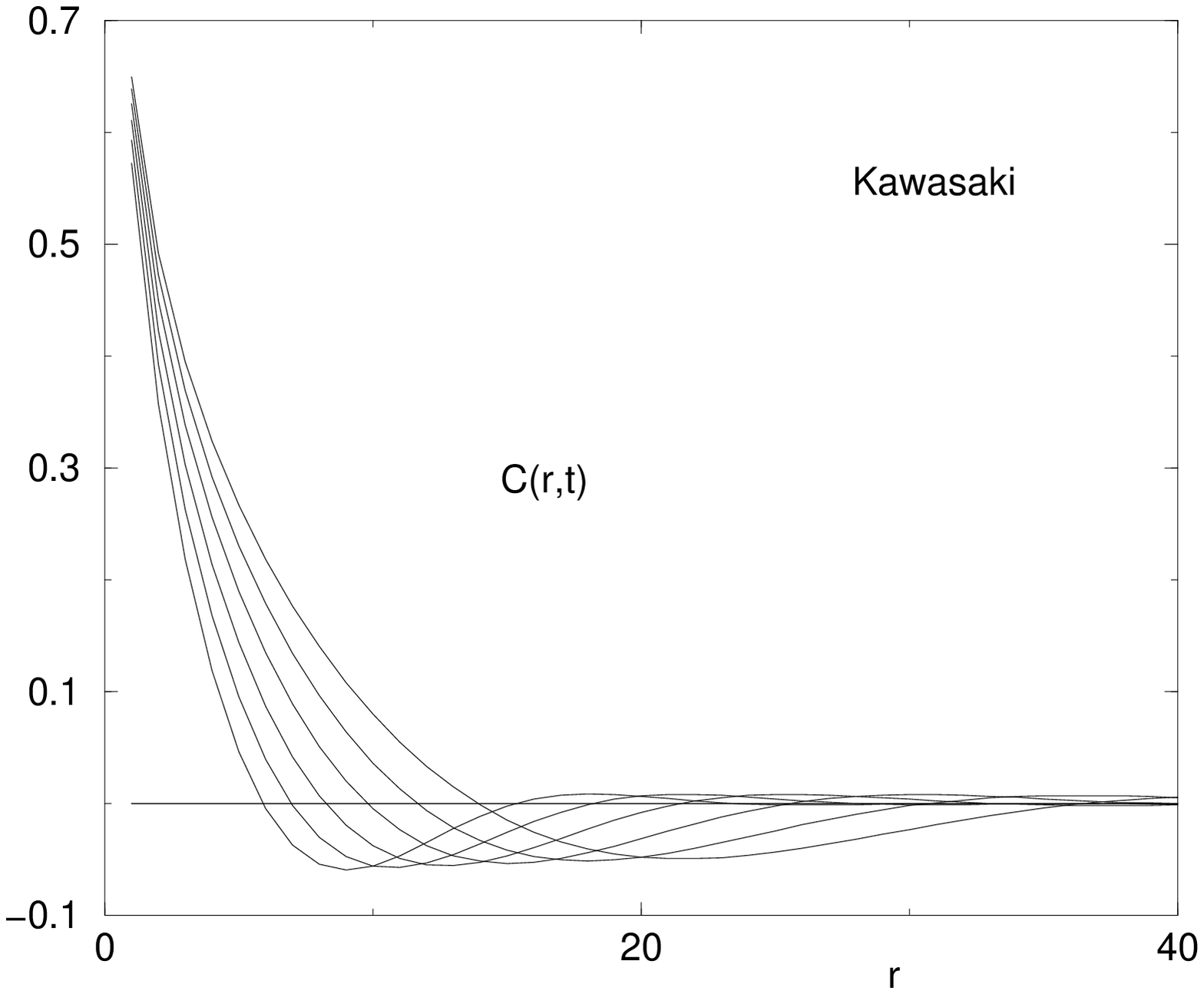}}}
\end{picture}
\caption{Correlation functions $C(r,t)$ for Kawasaki dynamics with 
the heat-bath algorithm. The lattice size is $L=512$.
From left to right, curves correspond to
the time $t=2000$, $4000$, $8000$,
$16000$, $32000$ and $64000$.
}
\label{f5}
\end{figure}

We regret that the data for Fig. \ref {f1} are obtained with
an incorrect updating scheme. The discussions and
conclusions based on Fig. \ref {f1} must be modified.
A correct version of Fig. \ref {f1} is displayed in Fig. \ref {f5}.
The scaling behavior described by Eq. (\ref {e10}) is observed.
The dynamic exponent $z$ depends slightly on the time $t$.
From our data, we could only conclude
that $z$ is $3.95(10)$. This is somewhat bigger
than the theoretical value $z=4-\eta=3.75$
calculated with the $\phi^4$ theory.
Probably there exist still corrections to scaling.
To perform simulations to larger time $t$'s and 
remove these corrections to scaling,
however, is not so easy since $z$ has a big value.
From the auto-correlation, we estimate the exponent
$\lambda=0.495(10)$. It is consistent with
the scaling relation $\lambda=d/z$ 
derived with the $\phi^4$ theory
(formally $\lambda=d/z-(4-z-\eta-\eta_0/2)/z=d/z+\eta_0/2z$, 
but $\eta_0=0$).

{\bf Acknowledgment}: The author thanks for
the visiting scholarship and
the hospitality from Institute of Theoretical Physics.
This work is supported in part by
DFG, Az. TR 300/3-1.

\end{document}